\documentclass{llncs}

\usepackage{graphicx}
\usepackage{amsmath}

\begin{document}

\title{A Software Reliability Model Based on a Geometric
       Sequence of Failure Rates\thanks{This
       research was partially supported by the DFG in the
       project 
       \emph{InTime}. 
       }}

\author{Stefan Wagner\inst{1} \and Helmut Fischer\inst{2}}

\institute{Institut f\"ur Informatik\\ Technische Universit\"{a}t M\"{u}nchen\\
Boltzmannstr.\ 3, 85748 Garching b.\ M\"unchen, Germany
\and Siemens AG\\ COM E QPP PSO\\ Hofmannstr.\ 51, 81379 M\"unchen,
Germany}

\maketitle

\begin{abstract}
Software reliability models are an important tool in quality management
and release planning. There is a large number of different models that
often exhibit strengths in different areas. This paper proposes a model that
is based on a geometric sequence (or progression) of the failure rates 
of faults. This
property of the failure process was observed in practice at Siemens among
others and led to the development of the proposed model. It is 
described in detail
and evaluated using standard criteria. Most importantly, the model 
performs constantly well over several projects in terms of its 
predictive validity.
\end{abstract}
 
\section{Introduction}

Software reliability engineering is an established area of software
engineering research and practice that is concerned with the improvement
and measurement of reliability. For the analysis typically stochastic software
reliability models are used. They model the failure process of the
software and use other software metrics or failure data as a basis
for parameter estimation. The models are able (1) to estimate the
current reliability and (2) to predict future failure behaviour.

There are already several established models. The most important ones
has been classified by Miller as exponential order statistic (EOS) models in
\cite{miller86}. He divided the models on the highest level into deterministic
and doubly stochastic EOS models arguing that the failure rates either have
a deterministic relationship or are again randomly distributed. For the
deterministic models, Miller presented several interesting special cases. The
well-known Jelinski-Moranda model \cite{jelinski72}, for example, has 
\emph{constant
rates}. He also stated that \emph{geometric rates} are
possible as documented by Nagel \cite{nagel82,nagel84}.

This geometric sequence (or progression) between failure rates of 
faults was also observed
in projects of the communication networks department of the Siemens AG.
In several older projects which were analysed, this relationship 
fitted well to the data. Therefore, a software reliability model based on
a geometric sequence of failure rates is proposed.

\paragraph{Problem.} The problem which software reliability engineering 
still faces is the need for accurate models for different environments
and projects. Detailed models with a geometric sequence of failure rates
have to our knowledge not been proposed so far.

\paragraph{Contribution.} We describe a detailed and practical software
reliability model that was motivated out of practical experience and
contains a geometric sequence of failure rates which was also suggested
by theoretical results. A detailed comparison shows that this model has
a constantly good performance over several projects, although other models
perform better in specific projects. Hence, we validated
the general assumption that a geometric sequence of failure rates
is a reasonable model for software.

\paragraph{Outline.} We first describe important aspects of
the model in Sec.~\ref{sec:description}. In Sec.~\ref{sec:evaluation} the
model is evaluated using several defined criteria, most importantly its
predictive validity in comparison with established models. 
We offer final conclusions in Sec.~\ref{sec:conclusions}. Related work
is cited where appropriate.

\section{Model Description}
\label{sec:description}

The core of the proposed model is a geometric sequence for the
failure rates of the faults. This section describes this and other assumptions 
in more detail, introduces the main equations and the time
component of the model and gives an example of how the parameters of the
model can be estimated.

\subsection{Assumptions}

The main theory behind this model is the ordering of the faults
that are present in the software based on their failure rates. The term
failure rate describes in this context the probability that an
existing fault will result in an erroneous behaviour of the system
during a defined time slot or while executing an average operation.
In essence, we assign each fault a time-dependent probability of failure
and combine those probabilities to the total failure intensity.
The ordering implies
that the fault with the highest probability of triggering a failure comes
first, then the fault with the second highest probability and so on.
The probabilities are then arranged on a logarithmic scale to attain
an uniform distribution of the points on the $x$-axis. 
The underlying assumption being that there are numerous faults with
low failure rates and only a small number of faults with high failure
rates. 
In principle, we assume an infinite number of faults because of imperfect
debugging and updates.

As mentioned above, the logarithmic scale distributes the data points in
approximately the same distance from each other.
Therefore, this distance is approximated by a constant factor between 
the probabilities.
Then we can use the following geometric sequence (or progression) for the 
calculation of the failure rates:
\begin{equation}
p_n = p_1 \cdot d^{(n-1)},
\end{equation}
where $p_n$ is the failure rate of the $n$-th fault, $p_1$ the failure
rate of the first fault, and $d$ is a project-specific parameter. It is
assumed that $d$ is an indicator for the complexity of a system that
may be related to the number of different branches in a program. In
past projects of Siemens $d$ was calculated to be between $0.92$ and $0.96$. 
The parameter $d$
is multiplied and not added because the distance is only constant on
a logarithmic scale.

The failure occurrence of a fault is assumed to be geometrically distributed.
Therefore, the probability
that a specific fault occurred by time $t$ is the following:
\begin{equation}
P(T_a \leq t) = F_a(t) = 1 - (1 - p_a)^t.
\label{eq:fischer_lambda}
\end{equation}
We denote with $T_a$ the random variable of the failure time of the fault $a$.

In summary, the model can be described as the sum of an infinite
number of geometrically distributed random variables with different
parameters which in turn are described by a geometric sequence.

\subsection{Equations}

The two equations that are typically used to describe a software reliability
model are the mean number of failures $\mu(t)$ and the failure intensity
$\lambda(t)$.
The mean value function needs to consider the expected value over the
indicator functions of the faults: 
\begin{equation}
\begin{array}{ll}
\mu(t) & = E(N(t))\\
     & = E\left (\sum_{i=a}^\infty{I_{[0,t]}(X_a)} \right )\\
     & = \sum_{a=1}^\infty{E(I_{[0,t]}(X_a))}\\
     & = \sum_{a=1}^\infty{P(X_a \leq t)}\\
     & = \sum_{a=1}^\infty{1 - (1 - p_a)^t}.
\end{array}
\label{eq:fischer_mean_new}
\end{equation}

\begin{figure}[h]
  \centering \includegraphics[width=.8\textwidth]{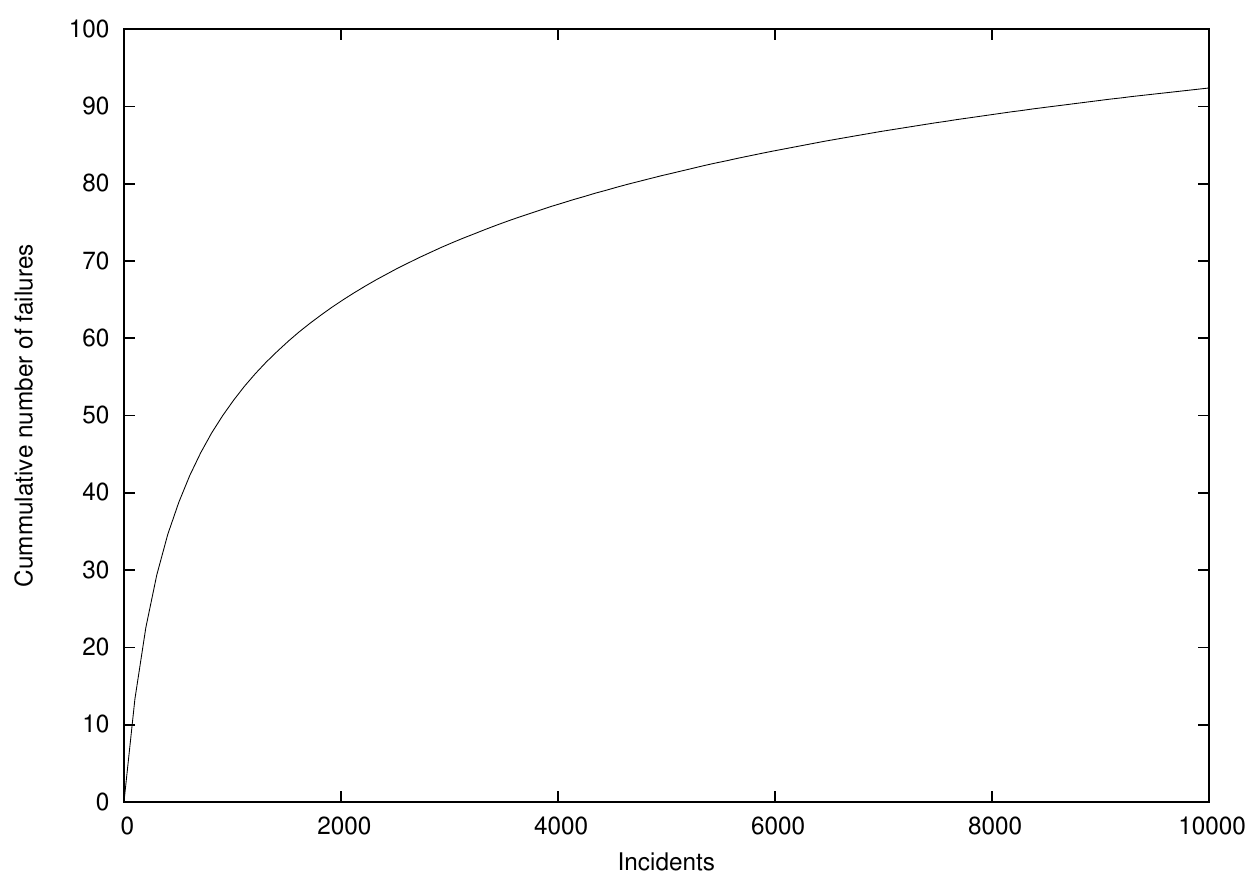}
  \caption{A typical distribution of the model} 
  \label{fig:typical_curve}
\end{figure}

This gives us a typical distribution as depicted in 
Fig.~\ref{fig:typical_curve}. Note that the distribution is actually
discrete which is not explicitly shown because of the high values
used on the $x$-axis.

We cannot differentiate the mean value equation directly
to get the failure intensity. However, we can use the probability
density function (pdf) of the geometric distribution to derive this
equation. The pdf of a single fault is
\begin{equation}
f(t) = p_a (1-p_a)^{t-1}.
\label{eq:failure_intensity}
\end{equation}
Therefore, to get the number of failures that occur at a certain point
in time $t$, we have to sum up the pdf's of all the faults:
\begin{equation}
\lambda(t) = \sum_{a=1}^\infty{p_a (1-p_a)^{t-1}}.
\end{equation}

An interesting quantity is typically the time that is needed to reach
a certain reliability level. Based on the failure
intensity objective that is anticipated for
the release, this can be derived using the equation for the failure
intensity. Rearranging Eq.~\ref{eq:failure_intensity} gives:
\begin{equation}
t = \frac{\ln \lambda}{\sum_{a=1}^\infty{p_a - p^2_a}} + 1.
\end{equation}
What we need, however, is the further required time $\Delta t$ to
determine the necessary length of the test or field trial. We denote the 
failure intensity
objective $\lambda_F$ and use the following equation to determine
$\Delta t$:
\begin{equation}
\Delta t = t_F - t = \frac{\ln \lambda_F - \ln \lambda}
                          {\sum_{a=1}^\infty{p_a - p^2_a}}
\end{equation}
Finally, the result needs to be converted into calendar time to
be able to give a date for the end of the test or field trial.

\subsection{Time Component}
\label{sec:time_component}

In the proposed model time is measured in incidents, each representing
a usage task of the system. To convert these incidents 
into calendar time it is necessary to introduce an
explicit time component. This contains
explicit means to convert
from one time format into another.

\begin{figure}[h]
  \centering \includegraphics[width=0.7\textwidth]{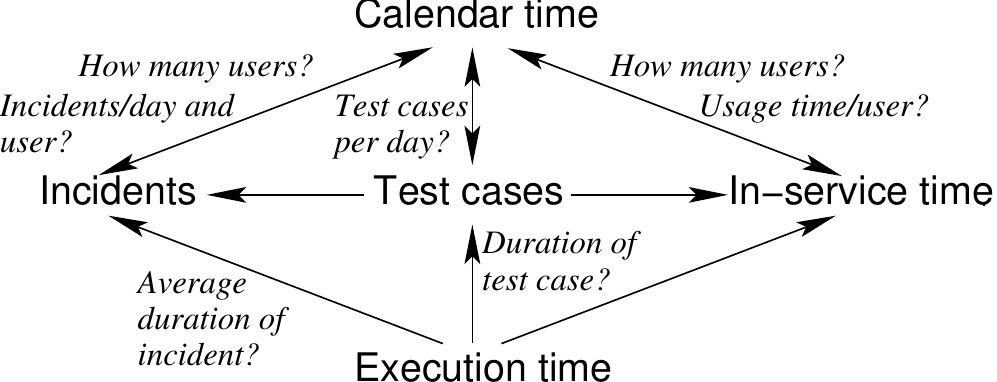}
  \caption{The possible relationships between different types of time}
  \label{fig:times}
\end{figure}

There are several possibilities
to handle time in reliability models. The preferable is to
use execution time directly. This, however, is often not possible. 
Subsequently, a suitable substitute must be found. With respect to
testing this could be the number of test cases, for the field use the
number of clients and so forth. Fig.~\ref{fig:times} shows the relationships
between different possible time types.

The first possibility is to use in-service time as a substitute. This
requires knowledge of the number of users and the average usage time per user.
Then the question arises how this relates to the test cases in system
testing. A first approximation is the average duration of a test case.
The number of incidents is, opposed to the in-service time, a more
task-oriented way to measure
time. The main advantage of using incidents, apart from the fact that they
are already in use at Siemens, is that in this way, we can obtain very
intuitive metrics, e.g., the average number of failures per incident. There
are usually some estimations of the number of incidents per client and 
data about the number of sold client licenses. 

However, the
question of the relation to test cases is also open. A first cut would be
to assume a test case is equal to an incident. A test case, however, has
more ``time value'' than one incident because it is generally directed
testing, i.e., cases with a high probability of failure
are preferred. In addition, a test case is usually unique in function
or parameter set while the normal use of a product often consists
of similar actions. When we do not follow the operational profile this
should be accounted for. A possible extension of the model is proposed in
\cite{wagner:tumi05} but needs further investigation.

\subsection{Parameter Estimation}
\label{sec:estimation}

There are two techniques for parameter determination currently in use. The
first is
prediction based on data from similar projects. This is useful for
planing purposes before failure data is available.

However, estimations should also be made during test, field trial, and operation
based on the sample data available so far. This is the approach
most reliability models use and it is also statistically most advisable
since the sample data comes from the population we actually want to analyse.
Techniques such as Maximum Likelihood estimation or Least
Squares estimation are used to fit the model to the actual
data. 

\paragraph{Maximum Likelihood.}

The Maximum Likelihood method essentially uses a likelihood function that
describes the probability of a certain number of failures occurring
up to a certain time. This function is filled with sample data and then
optimised to find the parameters with the maximum likelihood.

The problem with this is that the likelihood function of this model
gets extremely complicated. Essentially, we have an infinite number of
random variables that are geometrically distributed, but all with
different parameter $p$. Even if we constrain ourselves to a high
number $N$ of variables under consideration it still results in
a sum of $\tbinom{N}{x}$ different
products. This requires to sum up every possible permutation
in which $x$ failures have occurred up to time $t$. The number of
possibilities is $\tbinom{N}{x}$.
Each summand is a product of a permutation in which different faults
resulted in failures.
\begin{equation}
\begin{array}{ll}
L(p_1,d) = &
\prod_{i=1}^x{1 - (1 - p_i)^t} \cdot \prod_{i=x+1}^N{(1-p_i)^t} + \\
& \prod_{i=2}^{x+1}{1 - (1 - p_i)^t} \cdot \prod_{i=x+2}^N{(1-p_i)^t} \cdot
(1-p_1)^t + \\
& \prod_{i=3}^{x+2}{1 - (1 - p_i)^t} \cdot \prod_{i=x+3}^N{(1-p_i)^t} \cdot
\prod_{i=1}^2{(1-p_1)^t} + \\
& \ldots,
\end{array}
\end{equation}
where $p_i = p_1 d^{i-1}$.

An efficient method to maximise
this function has not been found.

\paragraph{Least Squares.}

For the Least Squares method an estimate of the failure
intensity is used and the relative error to the estimated failure
intensity from the model is minimised. We use the estimate of the
mean number of failures for this because it is the original part
of the model. Therefore, the square function to be minimised in
our case can be written as follows:
\begin{equation}
S(p_1,d) = \sum_{j=1}^m{[\ln r_j - \ln \mu(t_j;p_1,d)]^2},
\end{equation}
where $m$ is the number of measurement points, $r_j$ is the
measured value for the cumulated failures, and $t_j$ is the
time at measurement $j$.

This function is minimised using the simplex variant of Nelder
and Mead \cite{nelder65}. 
We found this method to be usable for our purpose.

\section{Evaluation}
\label{sec:evaluation}

We describe 
several criteria
that are used to assess the proposed model.

\subsection{Criteria}

The criteria that we use for the evaluation of the Fischer-Wagner model
are derived from Musa et al.\ \cite{musa87}. We assess according 
to five criteria,
four of which can mainly be applied theoretically, whereas one criterion
is based on practical applications of the models on real data. The first
criterion is the \emph{capability} of the model. It describes 
whether the model is able to yield
important quantities. The criterion \emph{quality of assumptions} is
used to assess the plausibility of the assumptions behind the model.
The cases in which the model can be used are evaluated with the criterion
\emph{applicability}. Furthermore, \emph{simplicity} is an important aspect
for the understandability of the model.
Finally, the \emph{predictive validity} is assessed by applying the
model to real failure data and comparing the deviation.

\subsection{Capability}

The main purpose of a reliability model is to aid managers and engineers
in planning and managing software projects by estimating useful quantities
about the software reliability and the reliability growth. Following
\cite{musa87} such quantities, in approximate order of importance, are

\begin{enumerate}
\item current reliability,
\item expected date of reaching a specified reliability,
\item human and computer resource and cost requirements related to
      the achievement of the objective.
\end{enumerate}

Furthermore, it is a valuable part of a reliability model if it can
predict quantities early in the development based on software metrics
and/or historical project data.

The model yields the current reliability as current failure intensity
and mean number of failures.
It is also able to
give predictions based on parameters from historical data.
Furthermore, the expected date of reaching a specified reliability
can be calculated. Human and computer resources are not explicitly
incorporated. There is an explicit concept of time but, it is not as
sophisticated as, for example, in the Musa-Okumoto model \cite{musa84}.

\subsection{Quality of Assumptions}

As far as possible, each assumption should be tested by real data. At
least it should be possible to argue for the plausibility of the
assumption based on theoretical knowledge and experience. Also the
clarity and explicitness of the assumptions are important.

The main assumption in the proposed model is that the failure rates of
the faults follow a geometric sequence. The intuition is that there are
many faults with low failure rates and only a small number of faults
with high failure rates. This is in accordance with software engineering
experience and supported by \cite{adams84}. Moreover, the geometric
sequence as relationship between different faults has been documented in a
NASA study \cite{nagel82,nagel84}.

Furthermore, an assumption is that the occurrence of a failure is
geometrically distributed. The geometric distribution fits because it
can describe independent events. We do not consider continuous
time but discrete incidents.

Finally, the infinite number of faults makes sense when considering
imperfect debugging, i.e., fault removal can introduce new faults or
the old faults are not completely removed.

\subsection{Applicability}

It is important for a general reliability model to be applicable to
software products in different domains and of different size. Also
varying project environments or life cycle phases should be feasible.
There are four special situations identified in \cite{musa87} that
should be possible to handle.

\begin{enumerate}
\item Software evolution
\item Classification of severity of failures into different categories
\item Ability to handle incomplete failure data with measurement
      uncertainties
\item Operation of the same program on computers of different performance
\end{enumerate}

All real applications of the proposed model have been in the
telecommunications
area. However, it was used for software
of various sizes and complexities. Moreover, during the evaluation of the
predictive validity we applied it also to other domains (see
Sec.~\ref{sec:validity}). In principle, the model can be used before and
during the field trial. Software evolution is hence not explicitly
incorporated. A classification of failures is possible but has not
been used so far. Moreover, the
performance
of computers is not a strong issue in this domain.

\subsection{Simplicity}

A model should be simple enough to be usable in real project
environments: it has to be simple to collect the
necessary data, easy to understand the concepts and assumptions, and
the model should be implementable in a tool.

While the concepts themselves are not difficult to understand, 
the model in total is rather complicated because it not only involves
failures but also faults. Furthermore, for all these faults the failure is
geometrically distributed but each with a different probability.

A main criticism is also that the assumed infinite number of faults
make the model difficult to handle. In practical applications
of the model and when building a tool, an upper bound of the number of
faults must be introduced to be able to calculate model values. This
actually introduces a third model parameter in some sense.

The two parameters, however, can be interpreted as direct measures
of the software.
The parameter $p_1$ is the failure probability of the most probable fault and 
$d$ can
be seen as a measure of system complexity.

\subsection{Predictive Validity}
\label{sec:validity}

The most important and ``hardest'' criterion for the evaluation of a
reliability model is its predictive validity. A model has to be a faithful
abstraction of the real failure process of the software and give valid
estimations and predictions of the reliability. For this we follow again
\cite{musa87} and use the \emph{number of failures approach}.

\subsubsection{Approach.}

We assume that there have been $q$ failures observed at the end of
test time (or field trial time) $t_q$. We use the failure data up to
$t_e (\leq t_q)$ to estimate the parameters of the mean number of failures
$\mu(t)$. The substitution of the estimates of the parameters yields the
estimate of the number of failures $\hat{\mu}(t_q)$. The estimate is
compared with the actual number at $q$. This procedure is repeated with
several $t_e$s.

For a comparison we can plot the relative error $(\hat{\mu}(t_q) - q) / q$
against the normalised test time $t_e/t_q$. The error will approach $0$ as
$t_e$ approaches $t_q$. If the points are positive, the model tends to
overestimate and accordingly underestimate if the points are negative. 
Numbers closer to $0$ imply a more
accurate prediction and, hence, a better model.

\subsubsection{Models for Comparison.}
As comparison models we apply four well-known models:
Musa basic, Musa-Okumoto, Littlewood-Verall, and NHPP. All these models
are implemented in the tool SMERFS \cite{farr93} that was used to
calculate the necessary predictions. We describe each model in more detail
in the following.

\paragraph{Musa basic.}
The Musa basic execution time model assumes that 
all faults are equally likely to occur, are independent
of each other and are actually observed. 
The execution times between failures are modelled as
piecewise exponentially distributed.
The intensity function is proportional  to the  number  of
faults remaining in the program and
the fault correction  rate is  proportional to the  failure
occurrence rate.

\paragraph{Musa-Okumoto.}
The Musa-Okumoto model, also called logarithmic Poisson execution time model,
was first described in \cite{musa84}. It also assumes that
all faults are equally likely to occur  and are independent
of each other.
The expected number of faults is a logarithmic function  of
time in this model, and
the failure intensity decreases exponentially with the expected 
failures experienced.
Finally, 
the  software will  experience an  infinite number of failures
in infinite time.

\paragraph{Littlewood-Verall Bayesian.} 
This model was proposed for the first time in \cite{littlewood73}.
The assumptions of the Littlewood-Verall Bayesian model are that
successive times between failures are independent random
variables each having an exponential distribution. The distribution
for the $i$-th failure has a mean of $1/\lambda(i)$.
The $\lambda(i)$s form a sequence of independent variables, each
having a gamma distribution with the parameters $\alpha$ and
$\phi(i)$. $\phi(i)$ has either the form:
$\beta(0) + \beta(1) \cdot i$ (linear)
or
$\beta(0) + \beta(1) \cdot i^2$ (quadratic). We used the quadratic
version of the model.

\paragraph{NHPP.}
Various models based on a non-homogeneous Poisson process are described
in \cite{pham00}. The particular model used
also assumes that
all faults are equally likely to occur  and are independent
of each other.
The cumulative number of faults detected at any time follows
a Poisson distribution with mean $m(t)$.  That  mean is
such that the expected number of faults in any small time
interval about $t$ is proportional to the number of undetected
faults at time $t$.
The mean is assumed to be a bounded non-decreasing function
with $m(t)$ approaching the expected total
number of faults to be detected
as the length of testing goes to infinity. It is possible
to use NHPP on time-between-failure data as well as failure counts. We
used the time-between-failure version in our evaluation.

\subsubsection{Data Sets.}
We apply the reliability models to several different sets of data
to compare the predictive validity. The detailed results for all of these
projects can be found in \cite{wagner:tumi05}. We describe only
the combined results in the following. The used data sets come (1) from
the \emph{The
Data \& Analysis Center for Software (DACS)} of the
US-American Department of Defence
and (2) from the telecommunication department of Siemens. The DACS data
has already been used in several evaluations of software reliability models.
Hence, this ensures the comparability of our results. In particular,
we used the projects 1, 6, and 40 and their failure data from system
tests measured in execution time. 

The Siemens data
gives additional insights and analysis of the applicability of the model
to these kind of projects. We mainly analyse two data sets containing
the failure data from the field trial of telecommunication software
and a web application. The Siemens data contains no execution time
but calendar time can be used as approximation because of 
constant usage during field trial.
All these projects come from different domains with
various sizes and requirements to ensure a representative evaluation.

\begin{figure}[h]
  \centering \includegraphics[width=.9\textwidth]{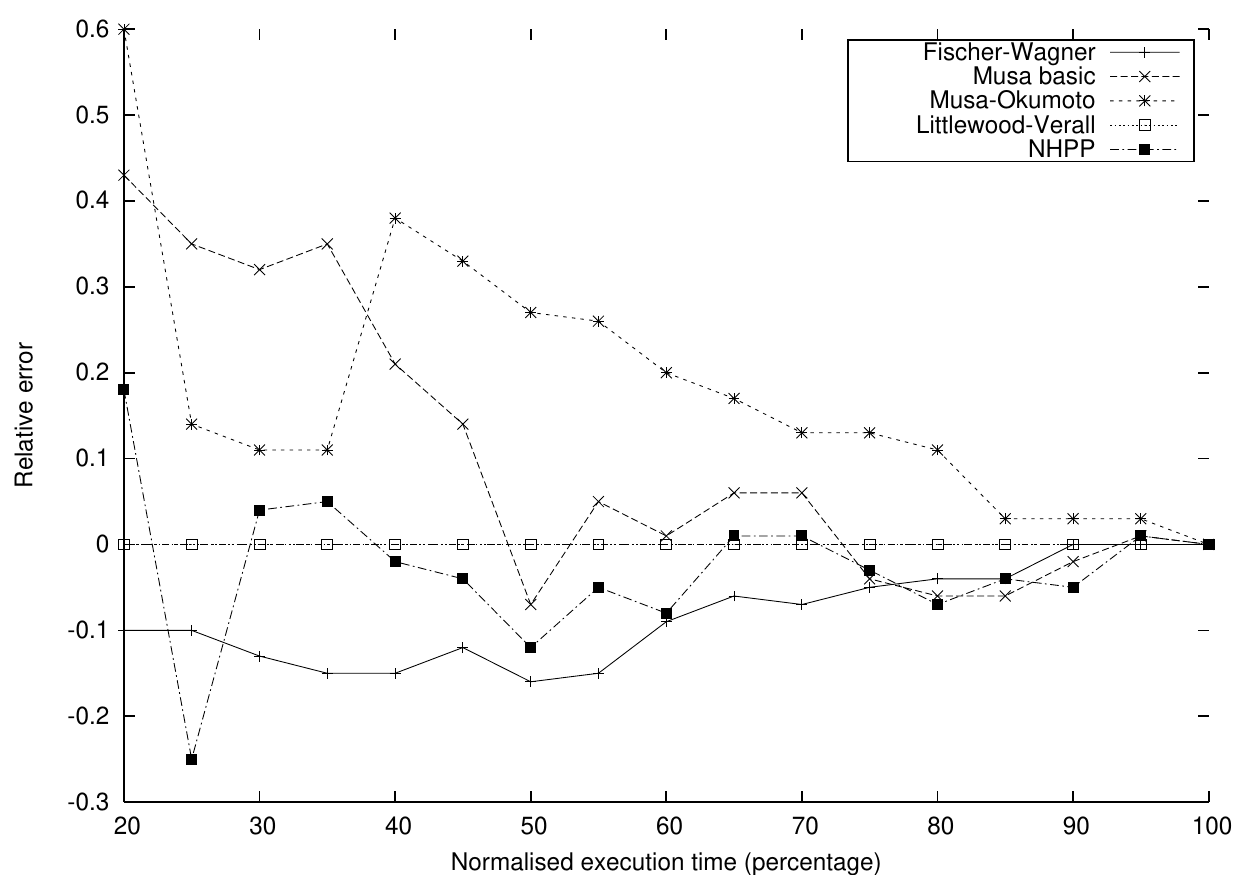}
  \caption{Median relative errors for the different models
           based on all analysed data sets}
  \label{fig:total}
\end{figure}

\subsubsection{Analysis and Interpretation.}
The usage of the number of failures approach for each project resulted
in different curves for the predictive validity over time. For a better
general comparison we combined the data into one plot which can be found
in Fig.~\ref{fig:total}. This combination is straight-forward as we only
considered relative time and relative errors. To avoid that strongly positive
and strongly negative values combined give very small errors we use
medians instead of average values. The plot shows that with regard to the
analysed projects the Littlewood-Verall model gives very accurate predictions,
also the NHPP and the proposed model are strong from early on.

However, for an accurate interpretation we have to note that the data of the
Littlewood-Verall model for one of the Siemens projects was not incorporated
into this comparison because its predictions were far off with a relative
error of about 6. Therefore, the model has an extremely good predictive
validity if it gives reasonable results but unacceptable predictions for
some projects. A similar argument can be made for the NHPP model which made
the weakest predictions for one of the DACS projects. The proposed
model cannot reach the validity of these models for particular projects,
but has a more constant performance over all projects. This is
important because it is difficult to determine which of the models
gives accurate predictions in the early stages of application since there
is only a small amount of data. Using the Littlewood-Verall or NHPP
model could lead to extremely bad predictions in some cases.

\section{Conclusions}
\label{sec:conclusions}

We conclude with a summary of our investigations and give some
directions for future work.

\subsubsection{Summary.}

We propose a software reliability model that is based on a geometric
series of the failure rates of faults. This basis is suggested from the
theory by Miller in \cite{miller86} as well as from practice in
Nagel et al.\ in \cite{nagel82,nagel84} and Siemens
projects.

The model has a state-of-the-art parameter determination approach
and a corresponding prototype implementation of it. Several data sets from
DACS and Siemens are used to evaluate the predictive validity of the
model in comparison to well-established models. We find that the
proposed model often has a similar predictive validity as the
comparison models and outperforms most of them. However, there is always
one of the models that performs better than ours. Nevertheless, we
are able to validate the assumption that a geometric sequence of
failure rates of faults is a reasonable model for software reliability.

\subsubsection{Future Work.}

The early estimation of the model parameters is always a problem
in reliability modelling. Therefore, we plan to evaluate the
correlation with other system parameters. For example
the parameter $d$ of the model is supposed to represent the
complexity of the system. Therefore, one or more complexity metrics
of the software code could be used for early prediction. This needs
extensive empirical analysis but could improve the estimation in the
early phases significantly.

Furthermore, a time component that also takes uncertainty into
account would be most accurate. The Musa basic and Musa-Okumoto models
were given such components (see \cite{musa87}). They model the usage
as a random process and give estimates about the corresponding calendar
time to an execution time.

Further applications with other data sets and comparison with other
types of prediction techniques, such as neural networks, are necessary
to evaluate the general applicability and predictive validity of the
proposed model.

Finally, we plan to use the model in an economics models for software quality
\cite{wagner:wosq05} and work further on a possibility to estimate
the test efficiency using the proposed model. Some early ideas
are presented in \cite{wagner:tumi05}.

\subsection*{Acknowledgements}

We are grateful to Christine Dietrich and Lothar Quoll for their help in
gathering the necessary data and to Sebastian Winter for useful comments
on the paper.

\bibliographystyle{plain}

\noindent \textcopyright Springer-Verlag. The final publication is available at \\ \url{http://link.springer.com/chapter/10.1007/11767077\_12}.

\end{document}